\documentclass[12pt]{article}
\usepackage{amsmath}
\usepackage{amssymb}
\usepackage{amsfonts}
\usepackage{theorem}
\usepackage{bbm}
\usepackage{color}
\usepackage[T1]{fontenc}
\usepackage{times}

\newtheorem{dfn}{Definition}[section]
\newtheorem{tw}[dfn]{Theorem}
\newtheorem{prop}[dfn]{Proposition}
\newtheorem{rem}[dfn]{Remark}
\newtheorem{ex}[dfn]{Example}
\newtheorem{lem}[dfn]{Lemma}

\newtheorem{cor}[dfn]{Corollary}

\usepackage{anysize}
\usepackage{verbatim}
\usepackage{array}
\usepackage{enumerate}
\usepackage{amssymb} % symbole AMS
 \usepackage{amsmath}
\usepackage{fancyhdr}
\usepackage{euscript}
\usepackage{latexsym}
\def\qed{\hbox{\hskip 6pt\vrule width6pt height7ptdepth1pt  \hskip1pt}\bigskip}

\author{Micha\l \ Barski  \\ \small  Faculty of Mathematics, Cardinal Stefan Wyszy\'nski University in Warsaw, Poland
\\
\small Faculty of Mathematics and Computer Science, University of Leipzig, Germany\\ \small{\it Michal.Barski@math.uni-leipzig.de}
 \bigskip \\
Jacek Jakubowski\\ \small Institute of Mathematics, University of
Warsaw, Warsaw,  Poland
\\ \small{\it J.Jakubowski@mimuw.edu.pl}
\small and \\
Jerzy Zabczyk\\ \small Institute of Mathematics, Polish Academy of
Sciences,
     Warsaw,  Poland
\\ \small{\it zabczyk@impan.pl}     
   }

\title{\bf On incompleteness of bond markets \\
with infinite number of random factors
\thanks{The authors express thanks to the anonymous reviewers for
their valuable comments which were helpful while improving the early
version of the paper.\newline Research supported by Polish KBN Grant
P03A 034 29 ,,Stochastic evolution equations driven by L\'evy
noise''} }
\date{}

\begin{document}
\maketitle
\begin{abstract}
The completeness of a bond market model with infinite number of
sources of randomness on a finite time interval in the
Heath-Jarrow-Morton framework is studied. It is proved that the
market  is not complete. A construction of a bounded contingent
claim, which can not be replicated, is provided.
\end{abstract}

\noindent
\begin{quote}


\end{quote}

\bigskip

\section{Introduction}

We investigate the completeness of a continuous time zero-coupon
bond market model in the Heath-Jarrow-Morton framework, see Heath,
Jarrow and Morton (1992). Let $P(t,T)$,  $0\leq
t,T\leq\bar{S}<\infty$ be the price at time $t$ of a bond paying $1$
at maturity $T$ and $\hat{P}(t,T)$ be the discounted price;
$\bar{S}$ denotes a finite time horizon of the model. The paper is
devoted to the problem of completeness of the market. We say that a
market is complete if an arbitrary bounded random variable can be
replicated by an admissible strategy. Thus we are concerned with a
problem of replicating for, each $S\leq\bar{S}$, contingent claims
depending on the information available up to time $S$. If
$S=\bar{S}$ we have a situation which occurs in practice, when a
trader hedges a payoff which depends on the bond's prices up to time
$\bar{S}$, uses bonds with maturities not exceeding $\bar{S}$.
Strategies of this type will be called {\it natural}.

The bond market we study is infinite in the sense that the price
process is a function-valued process. The concept of the portfolio
and the strategy may be formalized in many possible ways. We base
our approach on the stochastic integration theory with respect to
the Hilbert space-valued martingales, as presented in M\'etivier
(1982) and Da Prato and Zabczyk (1992).

The problem of completeness has been investigated by several
authors. Bj\"ork, Di Masi, Kabanov and Runggaldier (1997) regarded
the price process in the space of continuous functions and obtained
conditions for approximate completeness. Cylindrical integration
theory, due to Mikulevicius and Rozovskii (1998), Mikulevicius and
Rozovskii (1999), and signed measures as portfolios were used in De
Donno and Pratelli (2004). The problem of replicating with the help
of natural strategies was discussed and partially solved in Carmona
and Tehranchi (2004). It was shown in Carmona and Tehranchi (2004),
with the use of the Malliavin calculus, that each contingent claim
of a special form can be replicated. The problem was also considered
in Aihara and Bagchi (2005). The lack of completeness was shown in
Taflin (2005) however, for a non-standard definition of the set of
contingent claims, namely for $D_{0}:=\bigcap_{p>1}L^p(\Omega)$.

In the present paper the problem of completeness in the class of all
bounded random variables is studied. We consider the case when
strategies take values in the space $G^\ast$ - the dual of the
Sobolev space $G=H^{1}[0,\bar{S}]$. We prove that under some natural
conditions, the bond market model is not complete. The main
contribution of this paper is the result on incompleteness for
bounded contingent claims. Moreover, we provide a construction of a
bounded random variable which can not be replicated. As a corollary
we obtain incompleteness when the strategies take values in
$L^2[0,\bar{S}]$, compare Aihara and Bagchi (2005). Our results seem
to be in a contrast with Theorem $4.1$ in Aihara and Bagchi (2005)
which states that the bond market is complete (see our Remark
\ref{rem o bagchi}).
 %In the last section
At the end, we  comment on solvability of the equation
\eqref{repr-ident},  crucial for  the problem of market
completeness. It turns out that we can solve \eqref{repr-ident} in
the class of all integrable processes satisfying some natural
condition, but the solutions do not have a natural financial
interpretation.

\section{The model}
\subsection{Process of bond prices}
We consider a bond market with a finite time horizon $\bar{S}$
defined on a filtered probability space $(\Omega, {\cal F},({\cal
F}_t)_{t \in [0, \bar{S}]},\mathbb{P})$ satisfying usual
conditions. The filtration is generated by  a sequence of
independent standard Wiener processes  $W^i$, $i=1,2,\ldots$, i.e.
${\cal F}_t$ is generated by $W^i(s)$, where $i=1,2,\ldots$,
$s\leq t$. We interpret
 $W_t=(W^1(t), W^2(t),...)$ as
a cylindrical Wiener process $W$ in $l^2$. The dynamics of the
forward rate curve $f$ is given in the form
\begin{equation}\label{eq:wz-for}
df(t,T)=\alpha(t,T)dt+\sum_{i=1}^{\infty}\sigma^i(t,T)dW^i(t),\qquad
t\in[0,\bar{S}], \ T\in[0,\bar{S}],
\end{equation}
 where we put
 \begin{equation}
\label{eq:alpha=sigma=0}
 \alpha(t, T) = \sigma^i(t, T) = 0 \qquad \text{for} \ t \geq T \ \text{and} \ i=1,2, \ldots.
\end{equation} The formula
\eqref{eq:wz-for} defines a family of processes parametrized by a
continuous parameter $T\in[0,\bar{S}]$. For each $T\in[0,\bar{S}]$
the formula  describes the evolution of the forward rate on the
interval $[0,\bar{S}]$. Denote by $\sigma_t$ a linear operator from
$l^2$ into $H:=L^2[0,\bar{S}]$ given by the formula
\begin{gather*}
(\sigma_tu)(T):=\sum_{i=1}^{\infty}\sigma^i(t,T)u^i, \quad
u=(u^i)_{i=1}^{\infty}\in l^2, \quad t,T\in[0,\bar{S}],
\end{gather*}
and $\alpha_t(T)=\alpha(t,T)$. In this notation the formula
\eqref{eq:wz-for} has the following form:
\begin{gather}
 df_t=\alpha_tdt+\sigma_tdW_t,
\end{gather}
or equivalently, using a stochastic integral:
\begin{gather}
f_t=f_0+\int_{0}^{t}\alpha_sds+\int_{0}^{t}\sigma_sdW_s,\qquad
t\in[0,\bar{S}].
\end{gather}
\noindent For the concept of the stochastic integral in Hilbert
spaces see M\'etivier (1982), Da Prato and Zabczyk (1992).

\noindent The drift coefficient $\alpha$ is assumed to be a process
taking values in the space $H$, with the norm denoted by
$\mid\cdot\mid_H$, satisfying Bochner integrability condition:
\begin{equation}\label{warunek na alpha}
\int_{0}^{\bar{S}}\mid\alpha_t\mid_{H}dt=\int_{0}^{\bar{S}}\left(\int_{0}^{\bar{S}}\alpha^2(t,
T)dT\right )^{\frac{1}{2}}dt <\infty \qquad \mathbb{P}-a.s..
\end{equation}
Processes $(\sigma_t)$ is assumed to be a predictable process taking
values in the space of Hilbert-Schmidt operators $L_{HS}(l^2,H)$:
\begin{align}\label{war sigma}
\|\sigma_t\|^2_{L_{HS}(l^2,H)}=\sum_{i=1}^{\infty}\mid\sigma^i_t\mid^2_{H}=\sum_{i=1}^{\infty}
\left(\int_{0}^{\bar{S}} \sigma^i(t, T)^2dT\right)<\infty \qquad
\mathbb{P}-a.s.,
\end{align}
and to satisfy the integrability condition:
\begin{equation}\label{warunek na sigma}
\int_{0}^{\bar{S}}\|\sigma_t\|^2_{L_{HS}(l^2,H)}dt=
\sum_{i=1}^{\infty} \int_{0}^{\bar{S}}\left(\int_{0}^{\bar{S}}
\sigma^i(t, T)^2dT \right) dt<\infty \qquad \mathbb{P}-a.s..
\end{equation}
Conditions \eqref{warunek na alpha} and \eqref{warunek na sigma} are
necessary and sufficient for the forward rate curve process $(f_t)$
to be continuous in $H$.

\noindent The short rate process $r$ is defined by  $r(t):=f(t,t)$
and the evolution of the money held in the savings account is given
by the equation:
\begin{gather*}
dB(t)=r(t)B(t)dt.
\end{gather*}
The bond price $P$ is a process defined by the following formula:
\begin{gather}
P(t,T)=e^{-\int_t^T f(t,u)du} ,
\end{gather}
and the discounted bond price $\hat{P}$, due to
\eqref{eq:alpha=sigma=0}, satisfies:
\begin{equation}\label{wz wyjsciowy na zdysk cene oblig}
\hat{P}(t,T):=B^{-1}(t)P(t,T)=e^{-\int_0^T f(t,u)du}.
\end{equation}
Let $G:=H^1[0,\bar{S}]$ be the Hilbert space of absolutely
continuous functions with square integrable first derivative
equipped with the norm:
\begin{gather*}
\mid g \mid^2_G:=\mid g(0)\mid^2 +
\int_{0}^{\bar{S}}\Big(\frac{dg}{ds}(s)\Big)^2ds, \qquad g\in G.
\end{gather*}

\noindent Note that the process $\hat{P}$ takes values in $G$. In
fact, since:
\begin{gather*}
\frac{d}{dT}\ \hat{P}(t,T)=-\hat{P}(t,T)f(t,T),
\end{gather*}
we have:
\begin{align*}
\mid \hat{P}_t\mid^2_{G}&=\hat{P}(t,0)^2+\int_{0}^{\bar{S}}\Big(\hat{P}(t,T)f(t,T)\Big)^2dT\\
&\leq \hat{P}(t,0)^2+C(t)\int_{0}^{\bar{S}}f(t,T)^2dT<\infty,
\end{align*}
where $C(t):=\sup_{T\in[0,\bar{S}]}\hat{P}^2(t,T)$ is finite because
$\hat{P}(t,T)$ is a continuous function of \ $T$. We will also
assume that the model is arbitrage-free, in the sense that the
process $\hat{P}(\cdot,T)$ is a local martingale for every
$T\in[0,\bar{S}]$, see Delbaen and Schachermayer (1994), Delbaen and
Schachermayer (1998). This postulate is satisfied if and only if the
following $HJM$-condition holds, see e.g. Heath, Jarrow and Morton
(1992), Jakubowski and Zabczyk (2007):
\begin{gather}\label{war HJM}
\int_{t}^{T}\alpha(t,u)du= \frac{1}{2}\Big|
\int_{t}^{T}\sigma(t,u)du \Big|^2_{l^2}\quad \forall \
t,T\in[0,\bar{S}] .
\end{gather}
Differentiating \eqref{war HJM} with respect to $T$ gives the
following formula:
\begin{gather}\label{war HJM po zrozniczkowaniu}
\alpha(t,T)=\sum_{i=1}^{\infty}\sigma^i(t,T)\int_{t}^{T}\sigma^i(t,u)du\quad
\forall \ t,T\in[0,\bar{S}].
\end{gather}
As a consequence of \eqref{war HJM po zrozniczkowaniu} we can write
the following expression for the process $\hat{P}$:
\begin{align}\label{wz na zdyskontowana obligacje-HJM}
d\hat{P}(t,T)=\hat{P}(t,T) \Big(
\sum_{i=1}^{\infty}b^i(t,T)dW^i(t)\Big),
\end{align}
where
\begin{align} \label{wz na funkcje  b}
b^i(t,T):=-\int_{0}^{T}\sigma^i(t,u)du, \qquad t,T\in[0,\bar{S}].
\end{align}
Let $(\Gamma_t)$ be a stochastic process with values in the space of
linear operators from $l^2$ into $G$ given by the formula:
\begin{gather}\label{def operatora Gamma}
(\Gamma_tu)(T):=\hat{P}(t,T)\sum_{i=1}^{\infty}b^i(t,T)u^i, \quad
u\in l^2,\quad t,T\in[0,\bar{S}].
\end{gather}

\begin{prop}\label{lem o calkowalnosci Gamma}
For every $t\in[0,\bar{S}]$, \ $\Gamma_t$ given by \eqref{def
operatora Gamma} is a Hilbert-Schmidt operator form $l^2$ to $G$.
Moreover, with probability one
\begin{gather} \label{jj1}
\int_{0}^{\bar{S}}\|\Gamma_t\|^2_{L_{HS}(l^2,G)}dt<\infty.
\end{gather}
\end{prop}
{\bf Proof:} At the beginning we show an auxiliary estimation. By
\eqref{wz wyjsciowy na zdysk cene oblig} we have:
\begin{gather}
\mid\hat{P}(t,T)\mid\leq e^{\int_{0}^{\bar{S}}\mid f(t,u)\mid
du}\leq e^{(\int_{0}^{\bar{S}}\mid f(t,u)\mid^2 du)^{\frac{1}{2}} \
(\bar{S})^{\frac{1}{2}}}=e^{\mid f_t\mid_H \
(\bar{S})^{\frac{1}{2}}}.
\end{gather}
Since $(f_t)$ is a continuous process in $H$, we conclude that the
function $t\longrightarrow \mid f_t\mid_H$ is bounded as a
continuous function on $[0,\bar{S}]$, i.e. there exists a constant
$A>0$ such that:
\begin{gather}\label{ogr na norme f jednostajne}
\sup_{t\in[0,\bar{S}]}\mid f_t\mid_{H}\leq A.
\end{gather}
Thus for some $B>0$ we have:
\begin{gather}\label{jednostajne ogr P^}
\mid \hat{P}(t,T)\mid\leq B \quad \forall \ t,T\in[0,\bar{S}].
\end{gather}
For any $i=1,2,...$, we have:
\begin{align}\label{wz na pochodna iloczynu}\nonumber
\frac{d}{dT}\left(\hat{P}(t,T)b^i(t,T)\right)&=
\frac{d}{dT}\hat{P}(t,T) \ b^i(t,T)
+\hat{P}(t,T)\frac{d}{dT}b^i(t,T)\\&=f(t,T)\hat{P}(t,T)\int_{0}^{T}\sigma^i(t,u)du-\hat{P}(t,T)\sigma^i(t,T).
\end{align}
Using \eqref{ogr na norme f jednostajne}, \eqref{jednostajne ogr
P^}, \eqref{wz na pochodna iloczynu} we can estimate the $G$-norm of
$\hat{P}(t,\cdot)b^i(t,\cdot)$ by the $H$-norm of $\sigma^i_t$:
\begin{align}\label{oszacowanie normy iloczynu}
\nonumber
\mid\hat{P}&(t,\cdot)b^i(t,\cdot)\mid^2_{G}= \int_{0}^{\bar{S}}\hat{P}(t,T)^2\left(f(t,T)\int_{0}^{T}\sigma^i(t,u)du-\sigma^i(t,T)\right)^2dT\\
\nonumber &\leq B^2\int_{0}^{\bar{S}} 2
\left( \left(f(t,T)\int_{0}^{T}\sigma^i(t,u)du\right)^2+\sigma^i(t,T)^2 \right)dT\\
\nonumber &\leq B^2
\left(2\int_{0}^{\bar{S}}f(t,T)^2\left(T\int_{0}^{T}\sigma^i(t,u)^2du\right)dT+2\mid
\sigma^i_t\mid^2_{H}\right)\\
\nonumber &\leq B^2 \left(2\int_{0}^{\bar{S}}f(t,T)^2 \bar{S}\mid
\sigma^i_t\mid^2_{H} dT+2\mid
\sigma^i_t\mid^2_{H}\right)\\
\nonumber &\leq 2 B^2\mid \sigma^i_t\mid^2_{H}\left(
\bar{S}\int_{0}^{\bar{S}}f(t,T)^2 dT +1\right)=2 B^2\mid
\sigma^i_t\mid^2_{H}\left(
\bar{S}\mid f_t\mid^2_{H} +1\right)\\
&\leq 2B^2(\bar{S}A+1)\mid \sigma^i_t\mid^2_H.
\end{align}
Let $(e^i)_{i=1}^{\infty}$ be a standard basis in $l^2$. In virtue
of \eqref{warunek na sigma} and \eqref{oszacowanie normy iloczynu}
we obtain the desired estimation:
\begin{align*}
\int_{0}^{\bar{S}}\|&\Gamma_t\|^2_{L_{HS}(l^2,G)}dt=\int_{0}^{\bar{S}}\sum_{i=1}^{\infty}\mid\Gamma_te^i\mid^2_{G}dt=
\int_{0}^{\bar{S}}\sum_{i=1}^{\infty}\mid\hat{P}(t,\cdot)b^i(t,\cdot)\mid^2_{G}dt\\
&\leq2B^2(\bar{S}A+1)\int_{0}^{\bar{S}}\sum_{i=1}^{\infty}\mid\sigma^i_t\mid^2_Hdt\leq2B^2(\bar{S}A+1)\int_{0}^{\bar{S}}\|\sigma_t\|^2_{L_{HS}(l^2,H)}dt<\infty.
\end{align*}
\hfill\qed

\noindent As an immediate consequence we obtain:
\begin{cor}\label{prop o mart lok}
The process $\hat{P}$ of discounted bond prices is a $G$-valued
local martingale.
\end{cor}
Since \eqref{jj1} holds, the equation \eqref{wz na zdyskontowana
obligacje-HJM} can be written in the form, see M\'etivier (1982), Da
Prato and Zabczyk (1992):
\begin{gather}\label{wz na zdyskontowana obligacje-HJM-skrot}
\hat{P}_t=\hat{P}_0+\int_{0}^{t}\Gamma_sdW_s, \quad t\in[0,\bar{S}].
\end{gather}

\subsection{Portfolios and strategies}
\subsubsection{Trading strategies}
In general portfolios $\varphi$ are identified with linear
functionals acting on a space in which the price process lives. For
the bond market the following classes of portfolios are considered
in literature.

\begin{enumerate}[A)]
\item Portfolios  consisting of finite or infinite number of
bonds:
\begin{gather*}
\varphi =\sum_{i=1}^{m}\alpha_i\delta_{\{T_i\}},\quad
m\in\mathbb{N}\cup\infty, \ \alpha_i, T_i \in [0,\bar{S}], \
i=1,2,...,m, \ \sum_{i=1}^{\infty}\mid\alpha_i\mid<\infty.
\end{gather*}
\item $\varphi$ are finite signed measures on the interval
$[0,\bar{S}]$.
\item $\varphi$ are bounded functionals on the space
$G=H^1[0,\bar{S}]$, shortly $\varphi\in G^\ast$, where $G^\ast$
denotes the dual space.
\end{enumerate}

\noindent The class $(A)$ has an obvious interpretation. Some
justification for using portfolios as finite signed measures or as
elements of $G^\ast$ can be found in De Donno and Pratelli (2004) or
in Bj\"ork, Di Masi, Kabanov and Runggaldier (1997) and Bj\"ork,
Kabanov and Runggaldier (1997). Let us recall that the space
$G^\ast$ contains finite signed measures on the interval
$[0,\bar{S}]$.

\begin{dfn}
A trading strategy is any predictable process with values in some
fixed class in (A) - (C).
\end{dfn}

\begin{dfn}
The (discounted) wealth process $\hat{V}^{\varphi}$ corresponding to
$\varphi$ is given by:
\begin{gather*}
\hat{V}_t^{\varphi}=\hat{V}^{\varphi}_{0}+ \int_{0}^{t}<\varphi_s, d
\hat{P}_s>_{G^\ast,G} \ ,\qquad t\in[0,\bar{S}],
\end{gather*}
\end{dfn}

\noindent The concept of stochastic integral will be discussed now.

\noindent Since $\hat{P}$ is a $G$-valued local martingale of the
form \eqref{wz na zdyskontowana obligacje-HJM-skrot}, the class of
integrands, see M\'etivier (1982), consists of all $G^\ast$-valued
predictable processes $\varphi$ satisfying:
\begin{align}\label{Metivier 3a}
&\int_{0}^{\bar{S}}\mid\varphi_t(Q^{\frac{1}{2}}_t)\mid^2_{G^\ast}
dt<\infty,
\end{align}
where
\begin{gather} \label{defQ}
Q_t=\Gamma_t\Gamma^{'}_t, \qquad t\in[0,\bar{S}].
\end{gather}
In \eqref{defQ} $\Gamma_t^{'}$ is the conjugate of $\Gamma_t$, i.e.
for all $a\in G$ and $b\in l^2$ we have
$<\Gamma^{'}_ta,b>_{l^2}=<~a,\Gamma_tb>_{G}$.

\noindent We say that a predictable process $\varphi$ is $\hat{P}$
integrable if $\varphi$ satisfies \eqref{Metivier 3a}. Note that
$\varphi$ can be $\hat{P}$ integrable although $P(P_t\in
Dom\varphi_t)=0$, so $<\varphi_t,P_t>$ is not defined. Thus from
financial point of view it is natural to assume that $\varphi$ takes
values in $G^\ast$ since $P$ lives in $G$. The construction of the
stochastic integral in M\'etivier (1982) is developed for a square
integrable martingale, but it can be extended to local martingales
by the localization procedure. Moreover, identifying Hilbert space
$G$ with its dual, with $\tilde{\varphi}\in G^\ast$ corresponding to
$\varphi\in G$, we have
\begin{align*}
\mid\varphi(Q_t^{\frac{1}{2}})\mid^2_{G^\ast}&=<(\Gamma_t\Gamma_t^{'})^\frac{1}{2}\tilde{\varphi},(\Gamma_t\Gamma_t^{'})^\frac{1}{2}\tilde{\varphi}>_{G}
=\mid\Gamma^{'}_t\tilde{\varphi}\mid^2_{G}=\mid\Gamma^{\ast}_t\tilde{\varphi}\mid^2_{G^\ast}.
\end{align*}
So, the condition $\eqref{Metivier 3a}$  can be reformulated as:
\begin{gather}\label{calkowalnosc wzgl P^}
\int_{0}^{\bar{S}}\mid\Gamma_s^\ast\varphi_s\mid^2_{l^2}ds<\infty.
\end{gather}
Therefore, $\varphi$ is $(\hat{P}_t)$ integrable if
\eqref{calkowalnosc wzgl P^} holds and then
\begin{gather}\label{def calki wzgl P^}
\int_{0}^{t}<\varphi_s, d \hat{P}_s>_{G^\ast,G} \
=\int_{0}^{t}<\Gamma^\ast_s\varphi_s,dW_s>_{l^2,l^2} .
\end{gather}

\noindent We define a class of admissible strategies in a standard
way, see, for instance, Karatzas and Shreve (1998), Hunt and Kennedy
(2005) and Jarrow and Madan (1991).
\begin{dfn}\label{def admissible portfolio}
A trading strategy $\varphi$ is admissible if it is $(\hat{P}_t)$
integrable and if the process
\begin{gather*}
\int_{0}^{t}<\varphi_s,d\hat{P}_s>_{G^\ast,G}, \qquad
t\in[0,\bar{S}]
\end{gather*}
is a martingale. The class of all admissible strategies will be
denoted by $\mathcal{A}$.
\end{dfn}

\begin{ex}
\noindent\emph{Assume that $\varphi$ is a $G^\ast$-valued
predictable process.
\begin{enumerate}[1)]
\item If $\varphi$ satisfies the integrability condition:
\begin{gather*}
\mathbf{E}\left(\int_{0}^{\bar{S}}\mid
\Gamma_s^\ast\varphi_s\mid^2_{l^2}ds\right)^\frac{1}{2}<\infty,
\end{gather*}
then $\varphi\in\mathcal{A}$. Indeed, the Burkholder-Davies-Gundy
inequality implies that the integral
$\int_{0}^{\cdot}<\varphi_s,d\hat{P}_s>_{G^\ast,G}$ is a martingale.
\item If there exists a constant $K>0$ such that the following condition holds:
\begin{gather*}
\left|\int_{0}^{t}<\varphi_s,d\hat{P}_s>_{G^\ast,G}\right|<K,
\qquad \forall \ t\in[0,\bar{S}],
\end{gather*}
then $\varphi\in\mathcal{A}$, because a bounded local martingale is
a martingale.\hfill$\square$
\end{enumerate}}
\end{ex}

\section{Incompleteness}
\subsection{Incompleteness in general case}
We define the completeness of the market in a usual way.
\begin{dfn}
Let $S\leq \bar{S}$. The bond market is complete on $[0,S]$ if for
any $\mathcal{F}_{S}$-measurable, bounded random variable $\xi$
there exists an admissible strategy $\varphi$ and a constant $c$
such that
\begin{equation}\label{wz na reprezentacje xi}
\xi = c + \int_{0}^{S}<\varphi_t, d \hat{P}_t>_{G^\ast,G} .
\end{equation}
\end{dfn}
Now we state the main theorem of the paper.
\begin{tw}\label{tw o zup dla L2} The bond market is not complete
 on $[0,S]$ for any $S\leq \bar{S}$.
\end{tw}

\noindent For the proof we will need the following lemmas.
\begin{lem}\label{tw dla surjekcji} (Appendix B in
Da Prato and Zabczyk (1992)) Let $X,Y,Z$ be three Hilbert spaces and
$A:X\longrightarrow Z$, $B:Y\longrightarrow Z$ two linear bounded
operators. Then $Im A\subseteq Im B$ if and only if there exists
constant $c>0$ such that $\|A^\ast f\|\leq c \|B^\ast f\|$ for all
$f\in Z^\ast$.
\end{lem}
\begin{lem}\label{lem o reprezentacji zm los w l^2}
Let $\xi$ be a square integrable, $\mathcal{F}_S$ measurable random
variable such that $\mathbf{E}\xi=0.$ Then $\xi$ can be represented
in the following form:
\begin{gather*}
\xi=\int_{0}^{S}<\psi_s,dW_s>_{l^2,l^2},
\end{gather*}
where $\psi$ is a predictable, $l^2$-valued process satisfying
condition $\mathbf{E}\int_{0}^{S}\mid\psi_s\mid^2_{l^2} ds <\infty$.
Moreover, \
$\mathbf{E}\xi^2=\mathbf{E}\int_{0}^{S}\mid\psi_s\mid^2_{l^2} ds$.
\end{lem}
{\bf  Proof:} This result is true when $W$ is finite dimensional,
see Lemma 18.11 in Kallenberg (2001). The generalization can be
obtained by the following arguments. Let
$\mathcal{G}_n\subseteq\mathcal{F}$ be a $\sigma$-field generated by
\begin{gather*}
W_{n}(s)=(W^{1}(s),W^{2}(s),...,W^{n}(s)), \quad s\leq S,
\end{gather*}
 and let
$\xi_n:=\mathbf{E}(\xi\mid\mathcal{G}_n)$. Since $\xi\in
L^2(\Omega)$ and
$\mathcal{G}_n\uparrow\mathcal{F}=\bigcup_{n\geq1}\mathcal{G}_n$, so
by the classical convergence theorem for martingales, see Corollary
7.22 and Theorem.7.23 in Kallenberg (2001),
$\xi_n\longrightarrow\xi$ in $L^2(\Omega)$. On the other hand we
have:
\begin{gather*}
\xi_n=\int_{0}^{S}\varphi_n(s)dW_{n}(s).
\end{gather*}
Setting $\tilde{\varphi}^n_s:=(\varphi_n(s),0,...)\in l^2$ we see
that
\begin{gather*}
\xi_n=\int_{0}^{S}<\tilde{\varphi}^n_s,dW_s>_{l^2,l^2}.
\end{gather*}
Hence
$\mathbf{E}(\xi_m-\xi_n)^2=\mathbf{E}\int_{0}^{S}(\tilde{\varphi}^n_s-\tilde{\varphi}^m_s)^2ds\longrightarrow0$
with $m,n\longrightarrow\infty$ and therefore
$\{\tilde{\varphi}^n\}$ is a Cauchy sequence in
$F=L^2(\Omega\times[0,S],\mathcal{F}_S\otimes\mathcal{B}[0,S],\mathbb{P}\times\lambda;l^2)$.
By the completeness of $F$ there exists a predictable limit
$\varphi\in F$ of the sequence $(\tilde{\varphi}_n)$. Due to the
fact that
\begin{gather*}
\int_{0}^{S}<\tilde{\varphi}^n_s,dW_s>_{l^2,l^2}\longrightarrow\int_{0}^{S}<\varphi_s,dW_s>_{l^2,l^2}
\ \text{in} \ L^2(\Omega),
\end{gather*}
we have the required representation:
$\xi=\int_{0}^{S}<\varphi_s,dW_s>_{l^2,l^2}$.\hfill\qed

The next lemma states a uniqueness of random variables
representation in a class of admissible strategies. This property is
crucial for the method which is used in the proof of the main
result.

\begin{lem}\label{lem jednoznacznosc reprezentacji}
Let $x,y\in\mathbb{R}$. Assume that $\psi$ is integrable with
respect to $W$ and such that $\int_{0}^{t}\psi_s dW_s, \ t\in[0,S]$
is a martingale. If for $\varphi\in\mathcal{A}$ the following
condition
\begin{gather}\label{rownosc calek}
x+\int_{0}^{S}<\varphi_s,d\hat{P}_s>_{G^\ast,G}=y+\int_{0}^{S}<\psi_s,dW_s>_{l^2,l^2},
\end{gather}
is satisfied, then $x=y$ and $\Gamma^\ast_s\varphi_s=\psi_s$ a.s.
wrt. $P\otimes\lambda$  on $\Omega\times[0,S]$, where $\lambda$
denotes a Lebesgue measure.
\end{lem}
{\bf Proof:} Taking expectations of both sides in \eqref{rownosc
calek} we immediately obtain that $x=y$.\\
Thus the following condition is satisfied:
\begin{gather*}
\int_{0}^{S}<\Gamma^\ast_s\varphi_s-\psi_s,dW_s>_{l^2,l^2}=0.
\end{gather*}
By Lemma $10.15$ in Da Prato and Zabczyk (1992), there exists a
standard, one dimensional Wiener process $B$ for which we have:
\begin{gather*}
\int_{0}^{t}<\Gamma^\ast_s\varphi_s-\psi_s,dW_s>_{l^2,l^2}=
\int_{0}^{t}\mid\Gamma^\ast_s\varphi_s-\psi_s\mid_{l^2}dB(s)\qquad
\forall \ t\in[0,S].
\end{gather*}
Thus the process
$\int_{0}^{t}\mid\Gamma^\ast_s\varphi_s-\psi_s\mid_{l^2}dB(s)$ is a
martingale which is equal to zero at time $S$. So, it is equal to
zero for every $t\in[0,S]$. By the uniqueness of the martingale
representation, see Theorem 18.10 in Kallenberg (2001), we conclude
that the integrand must be zero which implies:
$\Gamma^\ast_s\varphi_s=\psi_s$, $P\otimes\lambda$ a.s.\\
\phantom{a} \hfill\qed

\noindent In order to find a strategy which replicates a square
integrable contingent claim $X$ one can use Lemma \ref{lem o
reprezentacji zm los w l^2} to represent $X$ as
$$X=EX+\int_{0}^{S}<\psi_s,dW_s>_{l^2,l^2}, $$ and then find an
admissible strategy $\varphi$ by solving the following {\it
structural equation}:
\begin{gather}\label{repr-ident}
\int_{0}^{S}<\varphi_s,d\hat{P}_s>_{G^\ast,G}=\int_{0}^{S}<\psi_s,dW_s>_{l^2,l^2}.
\end{gather}

\subsubsection{Proof of Theorem \ref{tw o zup dla L2}}

\noindent  Due to the definition of  the integral with respect to
$\hat{P}$, the set of all final portfolio values starting from zero
initial endowment has the following structure:
\begin{gather*}
\left\{\int_{0}^{S}<\varphi_t,d\hat{P}_t>_{G^\ast,G} \ : \ \varphi
\in \mathcal{A} \right\}=
\left\{\int_{0}^{S}<\Gamma^{\ast}_t\varphi_t,dW_t>_{l^2,l^2} \ : \
\varphi \in \mathcal{A} \right\}.
\end{gather*}
We will show that the operators $\Gamma^{\ast}_t, \ t\in[0,S]$, are
not surjective and we construct process $\psi$ taking values in the
space $l^2$ satisfying conditions:
\begin{enumerate}[1)]
\item $\psi$ is integrable wrt. $W$ and such that $\int_{0}^{t}<\psi_t,dW_t>_{l^2,l^2}, \ t\in[0,S]$, is a martingale,
\item $\psi_t\notin Im(\Gamma_t^\ast)$ with positive $\mathbb{P}\otimes\lambda$ measure,
\item the random variable
$\int_{0}^{S}<\psi_t,dW_t>_{l^2,l^2}$ is bounded.
\end{enumerate}

\noindent Then, in virtue of Lemma \ref{lem jednoznacznosc
reprezentacji}, for any $c\in\mathbb{R}$ and
$\varphi\in\mathcal{A}$ we have:
\begin{gather*}
\int_{0}^{S}<\psi_t,dW_t>_{l^2,l^2}\neq c+\int_{0}^{S}
<\Gamma^{\ast}_t\varphi_t,dW_t>_{l^2,l^2}.
\end{gather*}
Thus the integral  $\int_{0}^{S}<\psi_t,dW_t>_{l^2,l^2}$ is a
bounded random variable which can not be replicated.
\\

\noindent By Proposition \ref{lem o calkowalnosci Gamma} the
operator $\Gamma_t$ is a Hilbert-Schmidt operator for any
$t\in[0,S]$. Thus $\Gamma_t$ is compact, so is $\Gamma^\ast_t$. As a
compact operator with values in infinite dimensional Hilbert space,
$\Gamma^{\ast}_t$ is not surjective.

\noindent Let us consider the self-adjoint operator
$(\Gamma^{\ast}_t\Gamma_t)^{\frac{1}{2}}:l^2\longrightarrow l^2$
which is also compact. For any $u\in l^2$ we have:
\begin{align*}
\mid(\Gamma^{\ast}_t\Gamma_t)^{\frac{1}{2}}u \mid^2_{l^2}=
<(\Gamma^{\ast}_t\Gamma_t)^{\frac{1}{2}}u,(\Gamma^{\ast}_t\Gamma_t)^{\frac{1}{2}}u>_{l^2}
=<\Gamma^{\ast}_t\Gamma_tu,u>_{l^2}=<\Gamma_tu,\Gamma_tu>_{G}=\mid
\Gamma_tu\mid _{G}^2 \ ,
\end{align*}
so by Lemma \ref{tw dla surjekcji} it follows that
$Im(\Gamma^{\ast}_t)=Im((\Gamma^{\ast}_t\Gamma_t)^\frac{1}{2})$.\\
By Proposition $1.8$ in Da Prato and Zabczyk (1992) the operator
$(\Gamma^{\ast}_t\Gamma_t)^\frac{1}{2}$ can be represented by the
formula:
$$
(\Gamma^{\ast}_t\Gamma_t)^\frac{1}{2}=\sum_{i=1}^{\infty}\lambda_i(t)
\ \  g^i_t\otimes g^i_t,
$$
where $\lambda_i(t)$ is a random variable and $g^i_t$ is an $l^2$-
valued random variable for i=1,2,... . Here "$\otimes$" denotes the
linear operation: $(a\otimes b)h=a<b,h>$ for \ $a,b,h\in l^2$.
Moreover, $\lambda_i$ and $g^i$ are predictable as functions of
$(\omega,t)$ and $\lambda_i(t)\longrightarrow_{i\rightarrow\infty}
0$ by compactness of $(\Gamma^{\ast}_t\Gamma_t)^\frac{1}{2}$.

\noindent Our aim now is to construct the process $\tilde{\psi}$
where $\tilde{\psi}_t=(\tilde{\psi}^1(t),\tilde{\psi}^2(t),...)\in
l^2$ such that it is not of the form
$\sum_{i=1}^{\infty}\lambda_i(t)g^i_t<g^i_t,u>_{l^2}$ for any $u\in
l^2$. This process must thus satisfy:
\begin{align*}
&\sum_{i=1}^{\infty}\left(\frac{\tilde{\psi}^i(t)}{\lambda_i(t)}\right)^2=\infty,\\
&\sum_{i=1}^{\infty}(\tilde{\psi}^i{(t)})^2<\infty.
\end{align*}
Let us define the sequence $(i_k(t))_{k=1,2,...}$ in the following
way:
\begin{align*}
i_1(t):=\inf\left\{i: \frac{1}{\lambda_i(t)}\geq 1\right\},\\
i_{k+1}(t):=\inf\left\{i>i_k:\frac{1}{\lambda_i(t)}\geq k\right\}
\end{align*}
and put
\[
\tilde{\psi}^i(t)= \
\begin{cases}
 \ 0 \quad \text{if} \ i\neq i_k(t)\\
 \ \frac{1}{k} \quad \text{if} \ i=i_k(t).
\end{cases}
\]
Then we have
$\sum_{i=1}^{\infty}\left(\frac{\tilde{\psi}^i(t)}{\lambda_i(t)}\right)^2\geq
\sum_{k=1}^{\infty}\frac{1}{k^2} \ k^2=\infty$ and
$\sum_{i=1}^{\infty}\tilde{\psi}^i(t)^2=\sum_{k=1}^{\infty}\frac{1}{k^2}<\infty$,
so the process is bounded in $l^2$. It is also predictable since it
is obtained by measurable operations on predictable elements.
Thus the process $\tilde{\psi}$ is integrable with respect to $W$ and $\int_{0}^{\cdot} <\tilde{\psi},dW>_{l^2,l^2}$ is a martingale.\\
Now let us define a stopping time $\tau$ as:
\begin{gather*}
\tau:=\inf\{t>0:\Big|
\int_{0}^{t}<\tilde{\psi}_t,dW_t>_{l^2,l^2}\Big| \geq 1\}\wedge S.
\end{gather*}
Finally, we define the required process as:
\begin{gather*}
\psi_t:=\tilde{\psi}_t\mathbf{1}_{[0,\tau)}(t).
\end{gather*}
\hfill\qed

\begin{rem} \label{wn1}
From Theorem \ref{tw o zup dla L2} with $S=\bar{S}$ it follows that
the bond market is not complete on $[0,\bar{S}]$ if traders can use
natural strategies only.
\end{rem}

\subsection{Comments on incompleteness in $L^2[0,\bar{S}]$}

In the paper  Aihara and Bagchi (2005)  strategies $ \varphi$ with
values in $L^2[0,\bar{S}]$, satisfying
\begin{gather}\label{war z Bagchi}
\mathbf{E}\left(\int_{0}^{\bar{S}}\mid\varphi_t\mid^2_{L^2[0,\bar{S}]}dt\right)<\infty
\end{gather}
are considered (see formula (4.2) in Aihara and Bagchi (2005)), and
 absolutely continuous measures are identified with
their densities.
 With this definition of strategies it is shown that the market is
 complete (see  Theorem 4.1. in
Aihara and Bagchi (2005)). We have, however, the following
proposition.
\begin{prop}\label{prop do Bagchi}
There exists a model of the bond market in which all
$L^2[0,\bar{S}]$-valued processes $\varphi$ satisfying \eqref{war z
Bagchi} are admissible strategies.
\end{prop}
{\bf Proof:} We will construct the model by defining the volatility
coefficient. Let us assume that $\tilde{\sigma}$ satisfies the
following conditions:
\begin{align}
&0 \leq\tilde{\sigma}^i(t,T)\leq K \quad i=1,2,\ldots  , \
(t,T)\in[0,\bar{S}]\times[0,\bar{S}]  \  \text{for some} \ \ K>0,\label{ogr_na_sigme}\\
&\mid\tilde{\sigma}^i_t\mid_{H}^{2} \leq\frac{1}{i^2}\quad
i=1,2,\ldots, t\in[0,\bar{S}],\label{ogr_na_norme_sigmy}
\end{align}
and define a new operator $\sigma$ as
\begin{gather}\label{nowa_sigma}
\sigma^i(t,T) = \
\begin{cases}
\tilde{\sigma}^i(t,T) \ &\text{if} \quad
\sum_{i=1}^{\infty}\int_{0}^{t}\tilde{\sigma}^i(s,T)dW^i(s)\geq 0 ,
\\[2ex]
0  &\text{if} \quad
\sum_{i=1}^{\infty}\int_{0}^{t}\tilde{\sigma}^i(s,T)dW^i(s)< 0.
\end{cases}
\end{gather}
Let the coefficient $\alpha$ be given by the HJM condition
\eqref{war HJM po zrozniczkowaniu}:
\begin{gather}\label{alfa w Bagchim}
\alpha(t,T)=\sum_{i=1}^{\infty}\sigma^i(t,T)\int_{t}^{T}\sigma^i(t,s)ds.
\end{gather}
It follows from \eqref{ogr_na_sigme},\ \eqref{ogr_na_norme_sigmy},\
\eqref{nowa_sigma} and \eqref{alfa w Bagchim} \ that coefficients
$\alpha$ and $\sigma$ satisfy \eqref{warunek na alpha} and
\eqref{warunek na sigma}. Assume that the initial forward rate curve
is nonnegative: $f(0,T)\geq 0$ for $T\in[0,\bar{S}]$. Then
\begin{gather*}
f(t,T)=f(0,T)+\int_{0}^{t}\alpha(s,T)ds +
\sum_{i=1}^{\infty}\int_{0}^{t}\sigma^i(s,T)dW^i(s)\geq 0, \quad
(t,T)\in[0,\bar{S}]\times[0,\bar{S}]
\end{gather*}
and thus $\hat{P}(t,T)=e^{-\int_{0}^{T}f(t,u)du}\leq 1$. It follows
from the condition \eqref{ogr_na_norme_sigmy} that:
\begin{align*}
\sum_{i=1}^{\infty}\int_{0}^{\bar{S}}b^i(t,T)^2dT&=\sum_{i=1}^{\infty}\int_{0}^{\bar{S}}\left(\int_{0}^{T}\sigma^i(t,u)du\right)^2dT\\
&\leq\sum_{i=1}^{\infty}\int_{0}^{\bar{ S}}\left(
T\int_{0}^{T}\sigma^i(t,u)^2du \right)
dT\leq\bar{S}^2\sum_{i=1}^{\infty}\frac{1}{i^2} \qquad  \forall
t\in[0,\bar{S}].
\end{align*}
As a consequence we obtain the following inequalities:
\begin{align}\label{war wynikajacy z Bagchi}\nonumber
\mathbf{E}\left(\int_{0}^{\bar{S}}\mid
\Gamma^{\ast}_t\varphi_t\mid^2_{l^2}dt\right)&=\mathbf{E}\left(\int_{0}^{\bar{S}}\sum_{i=1}^{\infty}
\Big(\int_{0}^{\bar{S}}\varphi_t(T)\hat{P}(t,T)b^i(t,T)dT\Big)^2dt)\right)\\[2ex]\nonumber
&\leq\mathbf{E}\left(\int_{0}^{\bar{S}}\Big(
\int_{0}^{\bar{S}}\varphi_t(T)^2dT\sum_{i=1}^{\infty}\int_{0}^{\bar{S}}b^i(t,T)^2dT\Big)dt\right)\\[2ex]
&\leq\bar{S}^2\sum_{i=1}^{\infty}\frac{1}{i^2}\mathbf{E}\left(\int_{0}^{\bar{S}}\int_{0}^{\bar{S}}\varphi_t(T)^2dTdt\right)<\infty.
\end{align}
In view of \eqref{war wynikajacy z Bagchi} we conclude that for each
$\varphi$ satisfying \eqref{war z Bagchi} the process
$\int_{0}^{\cdot}<\varphi_t,d\hat{P}_t>_{G^\ast,G}$ is a martingale
and thus $\varphi$ is admissible.\hfill\qed

Therefore it follows from Proposition \ref{prop do Bagchi} and
Theorem \ref{tw o zup dla L2} that
\begin{cor}\label{cor do Bagchi}
There exists an incomplete bond market for which all
$L^2[0,\bar{S}]$-valued processes satisfying \eqref{war z Bagchi}
are admissible.
\end{cor}

\begin{rem}\label{rem o bagchi}
Due to relation $L^2[0,\bar{S}]=H\supseteq G=H^1[0,\bar{S}]$ we can
treat the process $(\hat{P}_t)$ as taking values in the space $H$.
The inclusion $H^\ast\subseteq G^\ast$ expresses the fact, that if
we admit $H$ as a state space for the discounted bonds prices, then
the investor can use smaller class of strategies and as a
consequence, in this case, the market is also incomplete. It follows
from Corollary \ref{cor do Bagchi} that Theorem 4.1 in Aihara and
Bagchi (2005) is false.
 It was pointed out by one of the reviewers
that the market considered in Aihara and Bagchi (2005) is
approximately complete but the limit passage, performed in the proof
of Th. 4.1 to get completeness, is not correct.
\end{rem}

\subsection{Comments on admissibility}

\noindent Notice that Lemma \ref{lem jednoznacznosc reprezentacji}
can be reformulated in the following way. If for the $W$ integrable
processes $\gamma,\psi$, such that the integrals
$\int<\gamma_s,dW_s>_{l^2,l^2}$ and $\int<\psi_s,dW_s>_{l^2,l^2}$
are martingales, we have:
\begin{gather*}
x+\int_{0}^{S}<\gamma_s,dW_s>_{l^2,l^2}=y+\int_{0}^{S}<\psi_s,dW_s>_{l^2,l^2}
\end{gather*}
for some $x,y\in\mathbb{R}$, then $x=y$ and $\gamma=\psi$. It turns
out that this assertion is not true if we assume only the existence
of the integrals or if we additionally assume that the integrals are
bounded from below. A counterexample in a one dimensional case which
we show is based on Example $8$, page 237 in Lipcer and Shiryaev
(2001).
\begin{ex}
\noindent\emph{Let us consider a one dimensional Wiener process $B$
on the interval $[0,1]$. The following stopping time:
\begin{gather*}
\tau:=\inf\{t\in[0,1]: B^2(t)+t=1\}
\end{gather*}
satisfies $P(0<\tau<1)=1$. The process
\begin{gather*}
X(t):=-\frac{2B(t)}{(1-t^2)}\mathbf{1}_{\{t\leq\tau\}}
\end{gather*}
is integrable with respect to $B$ because the following estimation
holds:
\begin{gather*}
\int_{0}^{1}X(s)^2ds=4\int_{0}^{\tau}\frac{B(s)^2}{(1-s)^4}ds<\infty.
\end{gather*}
Applying the It\^o formula to the process $\frac{B(t)^2}{(1-t)^2}$
we obtain:
\begin{gather*}
\int_{0}^{1}X(s)dB(s)-\frac{1}{2}X(s)^2ds=-1-2\int_{0}^{\tau}B(s)^2\left(\frac{1}{(1-s)^4}-\frac{1}{(1-s)^3}\right)ds<-1.
\end{gather*}
As a consequence, the Dol\'eans-Dade exponent $M=\mathcal{E}(X)$,
which is a local martingale, is not a martingale because
\begin{gather*}
\mathbf{E}(M(1))=\mathbf{E}\left(e^{\int_{0}^{1}X(s)dB(s)-\frac{1}{2}X(s)^2ds}\right)<e^{-1}<M(0)=1.
\end{gather*}
The random variable $M(1)$ satisfies estimation $0<M(1)<e^{-1}$ and
thus the application of the martingale representation theorem to the
square integrable martingale $\mathbf{E}[M(1)\mid\mathcal{F}_t]$
provides:
\begin{gather*}
M(1)=\mathbf{E}[M(1)]+\int_{0}^{1}\gamma(s)dB(s),
\end{gather*}
where $\mathbf{E}\int_{0}^{1}\gamma^2(s)ds<\infty$. On the other
hand the application of the martingale representation theorem to the
local martingale $M$ provides, see Theorem 18.10 in Kallenberg
(2001):
\begin{gather*}
M(1)=M(0)+\int_{0}^{1}\psi(s)dB(s)=1+\int_{0}^{1}\psi(s)dB(s),
\end{gather*}
where $P(\int_{0}^{1}\psi^2(s)ds<\infty)=1$. Moreover, $\psi$
satisfies condition $\mathbf{E}\int_{0}^{1}\psi^2(s)ds=\infty$
because $M$ is not a martingale.\\
Summarizing, we have two different representations of the same
bounded random variable $M(1)$, i.e.
\begin{gather*}
\mathbf{E}[M(1)]=x\neq y=1;\qquad \gamma\neq\psi.
\end{gather*}
Moreover, both representations are bounded from below by
zero.\hfill$\square$}
\end{ex}

\section{ Solvability of the structural equation}
As we can see the problem of market completeness is strictly
connected with the existence of a solution to the structural
equation \eqref{repr-ident}.  Theorem \ref{tw o zup dla L2} shows
that \eqref{repr-ident} may not have a solution in the class of
$G^\ast$- valued admissible strategies. However, the equation
\eqref{repr-ident} can be considered in the class of all processes
stochastically integrable with respect to $\hat {P}$. In the
definition of the stochastic integral we follow M\'etivier (1982),
see also Peszat and Zabczyk (2007). The class of integrands consists
of all predictable processes $\varphi$ taking values in the space of
linear but not necessarily continuous functionals on $G$, satisfying
the following conditions:
\begin{align}\label{Metivier 1}
&\text{Im} Q^{\frac{1}{2}}_t\subseteq \text{Dom}\varphi_t,\quad
\varphi_t(Q_t^{\frac{1}{2}})\in G^\ast, \quad
\forall t\in[0,S],\\
\label{Metivier 3}
&\int_{0}^{\bar{S}}\mid\varphi_t(Q^{\frac{1}{2}}_t)\mid^2_{G^\ast}dt<\infty.
\end{align}
\noindent The stochastic integral of $\varphi$ with respect to
$(\hat{P}_t)$ is denoted by $\int(\varphi_s,d\hat{P}_s)$. One can
show that if $\hat{P}$ is given by (\ref{wz na zdyskontowana obligacje-HJM-skrot}) then
\begin{gather} \label{calki-ogolne}
\int_{0}^{t}(\varphi_s,d\hat{P}_s)=\int_{0}^{t}< \Gamma_s^\ast \circ
\varphi_s,dW_s>_{l^2,l^2} , \qquad t\in[0,\bar{S}],
\end{gather}
where  $\Gamma_s^\ast \circ \varphi_s$ is defined by the formula
$\Gamma_s^\ast \circ \varphi_s(u)$ = $\varphi_s (\Gamma_s u), u\in
l^2.$

\noindent
  {\it If we enlarge the class of processes, in which we search
solutions of \eqref{repr-ident}, to all stochastically integrable
processes, then, under some assumptions, the structural equation
\eqref{repr-ident} has a solution, see the theorem and the example
below. Nevertheless, this solution does not have a natural
interpretation as a strategy and should be treated as a mathematical
idealization. }
\begin{tw}\label{tw o iniekcji}
If the operator $\Gamma$ is injective $\mathbb{P}\otimes\lambda$
a.s. then the equation \eqref{repr-ident} has a solution in the
class of integrable processes satisfying
\begin{gather} \label{jj2}
\mathbf{E}\left(\int_{0}^{\bar{S}}\mid\varphi_t(Q^{\frac{1}{2}}_t)\mid^2_{G^\ast}dt\right)<\infty.
\end{gather}
\end{tw}
{\bf Proof:} Fix  $\psi$ a predictable, $l^2$-valued process
satisfying condition
$\mathbf{E}\int_{0}^{\bar{S}}\mid\psi_s\mid^2_{l^2}~ds<~\infty$. We
will find an integrable process $\varphi$ satisfying \eqref{jj2}
such that
\begin{gather}\label{rownosc calek dla generalized}
 \int_{0}^{\bar{S}}<\psi_s,dW_s>_{l^2,l^2} = \int_{0}^{\bar{S}}(\varphi_s,
d\hat {P}_s) .
\end{gather}
 Since $\Gamma_t$ is injective, so
\begin{gather*}
\varphi_t(v):=<\psi_t,\Gamma_t^{-1}v>_{l^2}, \qquad \forall
v\in\text{Im}\Gamma_t,
\end{gather*}
is a well defined linear functional. The process
$(\varphi_t(\Gamma_t))$ is predictable and for any $u\in l^2$ we
have:
\begin{gather*}
<\psi_t,u>_{l^2}=<\psi_t,\Gamma_t^{-1}\Gamma_tu>_{l^2}=\varphi_t(\Gamma_tu).
\end{gather*}
Therefore
\begin{gather}\label{rownosc operatorow dla generalized}
\psi_t=\varphi_t(\Gamma_t), \qquad \forall t\in[0,\bar{S}],
\end{gather}
so by \eqref{calki-ogolne} the formula \eqref{rownosc calek dla
generalized} is satisfied. \hfill\qed

\noindent Now, we give an example of a bond market with
deterministic volatility (Gaussian HJM-model) in which the equation
\eqref{repr-ident} has a solution.

\begin{ex}\emph{
Let $\sigma^j, j=1,2,...$ be given by the formula:
\begin{gather*}
\sigma^j(t,T):=\gamma_j\sin\left(j\pi\Big(\frac{T-t}{\bar{S}-t}\Big)\vee
0\right), \quad 0\leq t,T\leq\bar{S},
\end{gather*}
where $\gamma_j>0$, and $\sum_{j=1}^{\infty}\gamma_i^2<\infty$.
Notice that for any $t\in[0,\bar{S}]$ the sequence
$(\sigma^j(t,\cdot))_j$ is an orthogonal system in $L^2[t,\bar{S}]$
and
\begin{gather*}
\mid\sigma^j(t,\cdot)\mid_{L^2[0,\bar{S}]}\leq\frac{\bar{S}}{2}\
\gamma^2_j.
\end{gather*}
Hence the process $(\sigma_t)$ satisfies \eqref{war sigma}. So the
deterministic process $(\Gamma_t)$ satisfies \eqref{jj1}. Therefore
the corresponding process $(\hat{P}_t)$ is a $G$-valued martingale.
To prove that in this case the equation \eqref{repr-ident} has a
solution, it is enough, by Theorem \ref{tw o iniekcji}, to show that
$\Gamma_t$ is injective for all $t\in[0,\bar{S}]$. To this end we
prove that if $\Gamma_tu=0$ for some $u\in l^2$, then u=0.
Differentiating
\begin{gather*}
\Gamma_tu(T)=-\sum_{j=1}^{\infty}\Big(\int_{0}^{T}\sigma^j(t,s)ds\Big)u^j
\end{gather*}
with respect to $T$, we see that
\begin{gather*}
\sum_{j=1}^{\infty}\sigma^j(t,s)u^j=0
\end{gather*}
in the sense of $L^2[t,\bar{S}]$. By orthogonality of the sequence
$(\sigma^j(t,\cdot))_j$ we obtain that $u^j=0$ for $j=1,2,...$,
hence $u=0$. So, $\Gamma$ is injective.}\hfill$\square$
\end{ex}

\begin{rem}
In general if $(\Gamma_t)$ is not injective
$\mathbb{P}\otimes\lambda$ a.s., then in a similar way as in the
proof of Theorem \ref{tw o zup dla L2} one can show that equation
\eqref{repr-ident} has no solution  even in the class of integrable
processes satisfying  \eqref{jj2}.
\end{rem}

\end{document}